\documentclass[preprint2]{emulateapj}
\usepackage{graphicx} \usepackage{rotating}
\usepackage{natbib} \bibpunct{(}{)}{;}{a}{}{,} 

\shorttitle{\Large Quark-Novae, Reionization, and Early R-Process Elements}
\shortauthors{Ouyed et al.}

\begin{document}

\title{Quark-Novae, cosmic reionization, and early r-process element production}

\author{Rachid Ouyed\altaffilmark{1,2}, Ralph
E. Pudritz\altaffilmark{2} and  Prashanth Jaikumar\altaffilmark{3}}

\altaffiltext{1}{Department of Physics and Astronomy, University of
Calgary, Calgary, Alberta  T2N 1N4, Canada; and Origins Institute
Senior Visiting Professor}
 
 \altaffiltext{2}{Origins Institute, ABB 241, McMaster
University, Hamilton, Ontario L8S 4M1, Canada}

 \altaffiltext{3}{Institute of Mathematical Sciences, CIT Campus,
Chennai, Tamil Nadu 600013, India and\\ Physics Division, Argonne
National Laboratory, Argonne, Illinois 60439, U.S.A.}
\email{ouyed@phas.ucalgary.ca}

\begin{abstract} We examine the case for Quark-Novae (QNe) as possible
sources for the reionization and early metal enrichment of the universe. 
Quark-Novae  are predicted to arise from the explosive collapse (and conversion) of sufficiently 
massive neutron stars into quark stars. 
 A Quark-Nova (QN) can occur over a range of time scales following
the supernova event.  For QNe that arise days to weeks after the supernovae,  
we show that dual-shock that arises as the QN ejecta encounter the
supernova ejecta can produce enough photons to reionize hydrogen in
 most of the  Inter-Galactic medium (IGM) by $z\sim 6$.  Such events can
explain the large optical depth $\tau_e\sim 0.1$ as measured by WMAP,
if the clumping factor, $C$,  of the material being ionized is smaller
 than 10. We suggest a way in which a normal initial mass function (IMF) for the oldest 
stars can be reconciled with a large optical depth as well as the mean
metallicity of the early IGM post reionization. We find that 
QN also make a contribution to r-process element abundances for 
atomic numbers $A \ge 130$.  We predict that the main 
cosmological signatures of Quark-Novae are the gamma-ray bursts that
announce their birth.  These will be clustered at redshifts in 
the range $z\sim 7$-$8$ in our model.

\end{abstract}  

\section{INTRODUCTION}

The end of the cosmic  ``dark age"  began with the production of
reionizing UV radiation from the first luminous objects.  Several
independent lines of evidence, including recent
WMAP results suggest that beginning at redshifts of about $z\sim 17$,
an inhomogenous and possibly non-monotonic decrease in the fraction
of neutral H and He in the IGM took place, and that this process was largely
completed by $z\sim 6$.  The nature of the first ionizing sources is
still mysterious - possible candidates for ``early'' sources (i.e.,
well before $z\sim 6$), include  metal-free Population III (Pop III) stars, 
high-$z$ dwarf galaxies, intermediate-mass black holes or sterile neutrinos.
Current models favour luminous objects that formed from collapsed
gas haloes of mass $M\sim 10^6M_{\odot}$.
     
Although quasars and active galactic nuclei (AGNs) are efficient
emitters of UV photons, results from the Sloan Digital Sky Survey
(SDSS) imply that they contribute much less to the ionizing radiation
than star-forming galaxies at high redshift (Fan et al. 2006). 
The first stars, owing to their metal-free composition ($Z$=0) which 
makes them a copious source of ionizing photons, are increasingly viewed 
as prime candidates for early reionization sources.  It has been argued
that a top-heavy IMF (40-100$M_{\odot}$) may be required to explain the large WMAP optical
depth and the ongoing overlap of HII regions at lower redshifts
$z\lesssim 9$ suggested by IGM temperatures inferred from the
Ly$\alpha$ forest (Theuns et al. 2002). 
Pop III stars derived from a top-heavy IMF
can emit as  many as $10^5$ ionizing photons per baryon during the
star's life (Tumlinson\&Shull 2000; Bromm et al. 2001; Schaerer
2002). A population  of very massive stars (VMS), that have been
linked to pair-instability  SNe, was advocated by Qian \& Wasserburg
(2002) to explain r-process element  abundances as well as that of C,
O and Si in some extremely metal-poor  stars, while other works
(Tumlinson et al.  2004;  Daigne et al. 2004;   Venkatesan \& Truran
2003; Fang\&Cen 2004) argue that an ordinary Type II-SN IMF can
explain these abundances just as well while also providing the large
optical depth $\tau_e$ measured by WMAP.  Furthermore, the escape
efficiency of UV photons required for  early reionization is smaller
for a top-heavy IMF than for the VMS,  making the former better
candidates for reionization sources.
 
While the high production of ionizing photons from a top-heavy
$Z$=$0$ stellar population is desirable, their short lifetimes (few
million years in the absence of significant mass loss) imply that
successive generations do not remain metal-free.  This creates
some major difficulties.  The issue here is that the IGM is
rapidly polluted in metals~\footnote{Analysis of quasar metal
absorption lines have shown that the IGM is enriched in heavy elements
from a level of $\sim
10^{-4}$ Z$_{\odot}$ at $z\sim 5.3$ to 
$\sim 10^{-3}$ Z$_{\odot}$ at $z\leq 4.3$   where Z$_{\odot}$ is the solar
metallicity (Songaila 2001). That such an abundance of heavy elements
is  observed commonly and uniformly in the low-density IGM where no
bright galaxies are found may be due to outflows from small
protogalaxies (Madau et al. 2001;  Mori et al. 2002), but prompt
enrichment by Pop III stars  provides a more natural solution for the
distribution of metagalactic heavy elements.}.  This ``negative
feedback'' mechanism (e.g. Sokasian et al. 2004; Ricotti \& Ostriker
2004) is a problem - it might result in the reneutralization of
the IGM, calling for another sub-epoch of reionization from a
different  stellar population in a metal-enriched IGM.  This may
necessitate double  or even multiple reionization (Wyithe \& Loeb
2003).  It may be possible to extend the Pop III lifetime by
``hiding''  most heavy elements in the collapse to black holes, with
added benefits  of additional accretion generated X-ray photons, but
this appears to  require extreme black hole formation rates and
smaller than normal metal yields from Pop II stars (see
Ricotti\&Ostriker 2004 for a discussion).

In this paper, we argue that explosions that result from 
 the conversion of neutron stars (born from progenitors with 
  mass in the $25M_{\odot}\le M\le 40M_{\odot}$ range) to quark stars
 - known as
Quark-Novae (QNe; Ouyed et al. 2002) -  may be important sources of reionizing UV photons.
This mechanism has the advantage of not leading to heavy chemical pollution of
the IGM or ISM, thereby alleviating the requirement  of producing several
generations of chemically pristine massive stars  early on in cosmic
star formation history.  The first QNe in the universe can occur very soon after
Pop III stars end their lives as Supernovae (SN) - in fact the time delay between
the SN and the subsequent QN is one of the main parameters of QN theory (Ouyed
 et al. 2007). The
ioinizing  photons are produced
when QN  shock waves overtake their progenitor SN shocks (Leahy\&Ouyed 2008).  Copious
energy releases can occur in this way if the QN delay is days to months
after the initial SN. QNe   can provide enough UV
photons to explain the large optical depth measured by WMAP. The
contribution to reionization from the main-sequence lifetime of the 
first stars is then reduced, which eliminates the need for a top-heavy IMF. The ``negative feedback'' problem of the increasing 
metallicity is mitigated by effectively decoupling metal-enrichment from 
reionization in the QN picture, without recourse to a large
formation rate of black holes. QNe also provide a local and 
prompt r-process enrichment for elements above $A\sim 130$, which can 
be related to the element abundance patterns in extremely metal-poor 
stars once chemical evolution studies are performed.

The properties of QN summarized above are the basis of 
our proposed modification of reionization and chemical
enrichment wherein: (i) the most massive stars ($M\ge
40M_{\odot}$) in a conventional or slightly top-heavy IMF collapse to
black-holes with a possible (small) contribution to reionization from
accretion but no contribution to metallicity while (ii) the
reionization is driven by intermediate-mass Pop III stars,  whose higher
mass members ($25M_{\odot}\le M\le 40M_{\odot}$) end up as
QNe, providing the bulk of reionzing photons and enriching
their environment in elements beyond $A\sim 130$, and (iii) low
mass Pop III stars ($8M_{\odot}\le M\le 25M_{\odot}$) end up
as type II SNe. A long-lived and metal-poor population of low-mass stars 
begins to emerge at the end of the  reionization epoch (Greif et al. (2008)). 
In our scenario, the dying out of the first heavy stars coincides with a 
peak in the QN rate and therefore a peak in ionizing radiation.

In this paper, we first outline the basic observational constraints
on reionization models (\S 2), followed by a description of QNe
 features in \S3 (the explosion, the compact remnant
  and the ejecta), and then derive the properties of
reionization by QNe (\S 4) as well as early r-process enrichment
(\S5). Our discussion and conclusions (\S 6) suggest observational
tests of this scenario.

\section{Observational Constraints on Reionization}

We first examine the constraints provided by the Gunn-Peterson test
and  the latest WMAP results.  Since even a tiny fraction of neutral H
($x_{\rm HI}\sim 10^{-3}$)  in the IGM is  sufficient to extinguish
the transmitted flux bluewards of the Ly$\alpha$  emission line from
distant Quasars, the onset of the Gunn-Peterson trough  (Gunn \&
Peterson 1965) in recent observations of Quasar emission spectra
(Schneider, Schmidt, \& Gunn 1991; see Djorgovski et al 2001 and references
therein) and the increased variance in the Gunn-Peterson optical depth
at $z\sim6$ shows that the IGM was transiting rapidly to a fully
reionized state by then.   This observation, although not a deep probe
of reionization history, places a floor on the emissivity of UV
photons required  to keep up with recombination during  the
reionization epoch  (Miralda-Escud\'e, Haehnet \& Rees 2000):

       \begin{equation}
\label{eq:floor} \dot{N}_{\rm ioni.} (z) = 10^{51.2}\ {\rm s}^{-1}\
{\rm Mpc}^{-3}\ \left(\frac{C}{30}\right) \left(\frac{1+z}{6}\right)^3
\left(\frac{\Omega_{\rm b}h^2}{0.02}\right)^2\ ,
       \end{equation}
        
 where $h= H_0/(100\ {\rm
km\ s}^{-1}\  {\rm Mpc}^{-1})$ is the scaled Hubble constant, and 
 $C$ is the clumping factor of the IGM related to small-scale ($\sim$ kpc)
  gas inhomogeneities caused by structure formation  (Shapiro\&Giroux 1987). 
We adopt a $\Lambda$CDM cosmology with $h= 0.72$, $\Omega_{\rm 0M}=0.26$ and
$\Omega_{\Lambda}=0.74$, $\Omega_{\rm b}h^2=0.023$ which are consistent 
with the 5-year WMAP data (Dunkley et al. 2009).

The WMAP results provide important constraints on the Thompson optical
depth at the time of reioinization.  Cosmic Microwave Background (CMB)
photons in the reionized epoch are  Thomson rescattered by free
electrons.  This damps the primary temperature anisotropy of the CMB
on scales smaller than the horizon size at rescattering, and generates
additonal polarization on large scales (tens of degrees). The WMAP
mission's measurements of the CMB polarization spectrum at low $l$
have quantified this effect through an  integrated Thomson optical
depth $\tau_e$. Recent analysis of 5-year WMAP data constrains it to be 
$\tau_{e}=0.087\pm 0.017$ for the best-fit $\Lambda$CDM cosmology 
(Dunkley et al. 2009); this is a much tightened value compared to 
the 3-year WMAP value $\tau_e=0.089\pm 0.030$ (Spergel et al. (2007)).
These results rule out prompt reionization at redshift $z\simeq
6$ at 3.5$\sigma$ and taken together with the varying Gunn-Peterson 
optical depth at $z\sim 6$, argue for an extended period of reionization
stretching back to $z\sim 20$. The smaller mass fraction in
collapsed halos at such high $z$ (e.g. Iliev et al. 2007 and
references therein) implies that a high emissivity of
ionizing photon per unit mass is required early in the
reionization epoch. This may be achieved with high formation
efficiency of metal-free massive Pop III stars, but as we show in
sections 3 and 4, our estimates for QN formation rates, the QN yield
in heavy elements and UV photons imply that QNe allow
for a more conventional star formation rate for Pop
III stars as a  viable alternative.

\section{The Quark-Nova}

If a neutron star  undergoes a deconfinement
phase transition to quark matter in its interior, it can undergo core
collapse, resulting in mass ejection of the outer crust (Takahara \& Sato 1986; Gentile et al. 1993; Fryer \& Woosley 1998;)  and the
formation of a stable strange quark star (a star
made of up, down and strange quarks; e.g. Itoh 1970; Bodmer 1971;
Witten 1984; Alcock et al. 1986, Bombaci et al. 2004).  Ouyed et
al. (2002) suggested that for sufficiently massive progenitors, the deconfined
cores physically separate from the overlaying hadronic layers giving
rise to powerful novae.  They  termed this the Quark-Nova scenario (see
 also Ker\"anen\&Ouyed 2003).

As outlined in Ker\"anen et al. 2005, the initial state for the QN is that
of a deleptonized neutron star with a {\it (u,d)} core.
In the quark-nova (QN) picture,  the {\it (u,d)}  quark
core of a hybrid star  that undergoes the phase transition to the {\it (u,d,s)}  quark phase, 
shrinks in a spherically symmetric fashion to a stable, more compact strange 
matter configuration faster than the overlaying material (the neutron-rich
hadronic envelope) can respond, leading to an effective core collapse.
  The two-step process, neutron to {\it (u,d)},
then {\it (u,d)} to {\it (u,d,s)} is crucial. 
In this scenario, the neutrinos come from weak reactions at the edge of the {\it (u,d)} 
core and can leak out easily into the surrounding cooler and deleptonized
envelope where they can deposit energy.  This
is significantly different from  phase conversion in a proto-neutron star stage where neutrino
transport is slower (of the order of seconds) because of the  hot and
lepton-rich matter.

A complete dynamical treatment of the QN, including neutrino
transport and full stellar evolution is only
just beginning to be explored (Sagert et al. 2009)
and preliminary results support a strong secondary explosion.  However,
we can make reliable estimates by adopting the viewpoint of numerous
previous authors  (Horvath \&
Benvenuto 1988;  Lugones et al. 1994; Anand et al. 1997; Ouyed et al. 2002; Ker\"anen et al. 2005; 
 Jaikumar et al. 2007; Sagert et al. 2009) who assume a
rapid adiabatic collapse on gravitational free-fall timescales ($\sim
10^{-4}$s) along with a conversion of the core to  ($u,d,s$) quark
matter on weak interaction timescales ($\sim 10^{-4}$s).  The
gravitational potential energy released in the collapse is converted
partly into latent heat of the (presumably) first-order phase
transition and  partly into outward propagating shock waves from the
supersonic motion of the ($u,d,s$) conversion front~\footnote{More
recent work, assuming realistic quark matter equations of state,
argues for strong deflagration (Drago et al. 2007), but this too may
be preceded by a compression shock (Lugones et al. 1994) in the
hadronic phase.}.  The temperature of the quark core thus formed rises
quickly to (10-20) MeV since the collapse is adiabatic rather than
isothermal (Gentile et al. 1993).

\subsection{The quark-nova compact remnant}

We assume that hot quark
matter in the color-flavor-locked (CFL) phase
is the true ground state of matter at high density.
 This is  a superconducting phase that 
 is energetically favored at extremely high densities and
low temperatures  (CFL;  Alford et al. 1999).
 In this phase u, d, and s quarks pair, forming a quark
condensate (a superfluid) that is antisymmetric in color and flavor indices.
  This state is reached by the QN compact remnant as it cools
  below a few tens of MeV (see Appendix A for a discussion
  on CFL and contamination of the universe).  
 
 Mannarelli et al. (2008) argue that CFL stars are unlikely to constitute a significant number of puslars since the r-mode is undamped for frequencies above 1Hz, due to the weak damping effect of mutual friction. However, this conclusion is derived for low temperatures $T\lesssim 0.01$MeV, while the temperature for the quark-nova is tens of MeV. As shown in Jaikumar et al. (2008), bulk viscosity from Kaons in the CFL phase is effective in damping out the r-mode at $T\geq 1$MeV, allowing the star to spin rapidly. Thus, the r-mode instability argument does not contradict the quark-nova mechanism (and the related CFL fireball; e.g. Vogt et al. 2004). However, it is clear that following the cooling phase of the quark-nova, the temperature will fall below 0.01 MeV. While mutual friction is too weak in this regime, other sources of viscosity may be important, such as shear viscosity from electromagnetic processes (it was shown in Manuel et al. (2005) that shear viscosity from phonons, though large, is inconsequential since it violates the basic hydrodynamic assumptions at such low temperatures). Therefore, we cannot comment on the ultimate fate of the CFL star below 0.01 MeV, but the quark-nova mechanism, which operates at much larger temperature, is not in obvious conflict with the r-mode instability.

\subsection{The quark-nova ejecta}

The gravitational potential energy
released (plus latent heat of
phase transition) during this event is converted partly into internal energy  and partly into outward propagating shock waves
which impart kinetic energy to the material that eventually forms the
ejecta.  Unlike Supernovae,
neutrino-driven mass ejection in Quark-novae is not feasible, as
neutrinos are trapped inside a hot and  dense expanding quark core,
once it grows to more than $\sim$2~km (Ker\"anen et al. 2005). Mass ejection due to core
bounce is also unlikely unless the quark core is very compact (1-2)km.
A more promising alternative is mass ejection from an expanding
thermal fireball that is a direct consequence of dense, hot quark
matter in the color-flavor-locked (CFL) phase (Jaikumar et al. 2002;
Vogt et al.  2004; Ouyed et al. 2005). Depending on the conversion
efficiency of photon  energy to kinetic energy of the ejecta, up to
$10^{-2}M_{\odot}$  of neutron-rich material can be ejected at nearly
relativistic speeds (Ouyed\&Leahy 2009).  Thus,  QNe ejecta  decompress material
from  the outer layers of exploding neutron stars. This is important
for the operation of the r-process (Jaikumar et al. 2007).

\subsection{Dual-shock quark-novae}

If  a QN occurs a few days to weeks
(Yasutake et al. 2005;  Staff et al. 2006) after the  Supernova,  the
QN ejecta propagating into the  Supernova envelope rapidly
sweep up enough mass to become sub-relativistic.  The collision of the
two ejecta sets up a blast wave that propagates outward and
re-energizes the Supernova ejecta. Part of the reshocked ejecta's
energy  is then radiated away in photons. This model for a
``dual-shock'' QN (hereafter dsQN) has been successfully applied to the
observed light-curve of the most energetic Supernovae (eq. SN2006gy,
SN2005ap, SN2005gj; see Leahy\&Ouyed 2008). It is this radiation
coming from the reshocked ejecta  which we wish to evaluate as a
potential reionizing source,  keeping in mind that such 
 dsQNe may be fairly frequent  (discussed below).

\section{Quark-Novae  and Reionization}   
  
 \subsection{The Ionization Efficiency}

We estimate the frequency of dsQNe, denoted by $f_{\rm dsQN}$, by assuming a
conventional IMF for massive stars (Scalo 1986),  an average Galaxy
density $n_g(z)$=$n_g(0)(1+z)^3$, and an  average Type II supernova
rate $f_{SN}(z)$ per year per Galaxy. For purpose of
estimation, we take $(1+z)\sim 10$ as typical of the reionization
epoch, $n_g(0)$=$0.02$/(Mpc)$^3$ and $f_{SN}(z)\sim 1$ yr$^{-1}$
Galaxy$^{-1}$. We find  $f_{\rm dsQN}=0.1x_{\rm
QN}f_{SN}n_g(0)(10)^3\sim 0.2$ yr$^{-1}$ Mpc$^{-3}$ where $x_{\rm QN}\sim 0.1$ 
is the fraction of stars in the 25-40$M_{\odot}$ range relative to all 
supernovae (8-40$M_{\odot}$) in the Scalo IMF. These stars are expected to 
undergo a Quark-Nova ~\footnote{ Staff et al. (2006) estimated that 1 out of a 1000 neutron stars could undergo the deconfinement transition during a Hubble time from spin-down alone. Mass fallback is more efficient at rapid conversion and upto 1 in 10 neutron stars with progenitors above 25$M_{\odot}$ can become quark stars.  Although this rate  may seem
high, Leahy \& Ouyed (2007) have argued that this is consistent with
the inferred birth rate of AXPs and SGRs (Gil\&Heyl
2007), which are  quark stars in their model.} with delay time between the supernova and the quark-nova 
decreasing with increasing mass (we assume conversion due to increased 
fall-back for more massive stars; see also Appendix B). The additional pre-factor of 0.1 is 
the fraction of quark-novae that can support an efficient dual shock. 
As argued in Leahy \& Ouyed (2008), to be consistent 
with the fraction of superluminous supernovae such as SN2006gy, the 
fraction of quark-novae which happen fast enough for the dual shock to 
be effective is typically about 0.1. Galaxy mergers or a longer 
ionization history can easily increase $f_{\rm dsQN}\sim$ 10  yr$^{-1}$ 
Mpc$^{-3}$. We choose $f_{\rm dsQN} \sim 1$ yr$^{-1}$ Mpc$^{-3}$ as an average 
frequency of QN events during the reionization epoch.
 
To determine the number of ionizing photons, we note that the
time-dependent luminosity of a dsQN is given by (Leahy\&Ouyed 2008)

\begin{equation}
\label{eq:luminosity} L_{\rm SN}(t)  =  c_{\rm v} \Delta T_{\rm core}
n_{\rm ejec.}  4\pi R_{\rm phot.}(t)^2 \frac{d D(t)}{d t}\ ,
\end{equation}

\noindent
where $c_{\rm v}\sim (3/2) k_{\rm B}$ is the specific heat of the
ejecta, $\Delta T_{\rm core}\sim T_{\rm core}$ is the core temperature
of the ejecta, $n_{\rm ejec}$ is the number density of the ejecta,
$R_{\rm phot.}(t)$ is the photospheric radius and $D(t)$ is the photon
diffusion length. This yields an integrated luminosity of
~$10^{51}$ergs,  in agreement with observations of SN2006gy, for the
parameter choice  $R_{\rm phot.}(t)\sim 3\times 10^{10}$km
~\footnote{$R_{\rm phot.}(t)$=$R_0+v_{SN}t$  with $R_0$ being the
radius of the progenitor star and $v_{SN}$ being the speed of the
shocked material.}. Furthermore, during time when the ejecta are
optically thick (tens of days for the most energetic Supernovae), the
luminosity can be approximated by $L_{\rm SN}(t)$=$\sigma 4\pi R_{\rm
phot}(t)^2T^4$ with $T\approx 10^4$K. This is close to the observed
spectral peak of  SN2006gy, in the first tens of days( Smith et al. (2007)). 
For metal-free Pop  III progenitors however, we expect the peak to lie in 
the UV region  ($T_{\rm pk}\approx (2-4)\times 10^4$K). The number of
ionizing photons  is then

\begin{equation}
\label{eq:ionizephotons} N_{\rm
ion}\approx\frac{N_0}{2.4}\int_{13.6/T_{\rm pk}({\rm eV})}^{\infty}
\frac{dx~x^2}{{\rm e}^x-1}
\end{equation}

\noindent
where $N_0$=$0.244(4\pi~R_{\rm phot}^3/3)T_{\rm pk}^3$ is the total
number of radiated photons. We find $N_0$=$2\times 10^{61}$-$2\times
10^{62}$ while  $N_{\rm ion}\approx 3\times 10^{59}$-$4\times 10^{61}$
for the lower and  upper limit of $T_{\rm pk}$ respectively. Since we
have neglected the  ionizing photon flux after the supernova ejecta
becomes transparent, our  estimate is a very conservative one. The
exact number of ionizing photons is very sensitive to $T_{\rm pk}$ and
$R_{\rm phot}(t)$, so we choose $R_{\rm phot}(t)$=3$\times 10^{15}$cm,
$T_{\rm pk}$=$3\times 10^4$K, giving  $N_{\rm ion}$=$5\times 10^{60}$
as a reasonable estimate (see also Appendix C). Then, the  corresponding emissivity of UV
ionizing photons is conveniently expressed as

        \begin{eqnarray} \dot{N}_{\rm ioni.,QNe}&\approx&  1.5\times
10^{53} \ {\rm s}^{-1}\  {\rm Mpc}^{-3}\\\nonumber &\times&
\left(\frac{f_{\rm dsQN}}{1\ {\rm yr}^{-1}\ {\rm Mpc}^{-3}}\right)
\left(\frac{N_{\rm ion}}{5\times 10^{60}\ {\rm photons}}\right)\ .
        \end{eqnarray}

\noindent
which exceeds the floor on the emissivity of ionizing photons from
Equation~(\ref{eq:floor}), even for the lower limit on  $N_{\rm
ion}\sim 3\times 10^{59}$.

It follows that the total number of ionizing photons {\it per baryon}
produced in dsQNe during the epoch  $\delta t_{\rm
re}$  (between $z\sim 17$ and $z\sim 6$) is

       \begin{eqnarray}
          \label{eq:photons} f_{\rm re} &\sim& 2\left(\frac{N_{\rm
ion}}{5\times 10^{60}}\right) \left(\frac{f_{\rm dsQN}}{1 {\rm
yr}^{-1}\ {\rm Mpc}^{-3}}\right) \\\nonumber
&\times&\left(\frac{\delta t_{\rm re}}{1\ {\rm Gyr}}\right)
\left(\frac{R_{\rm uni.}}{1\ {\rm Gpc}}\right)^3\left(\frac{N_{\rm
univ.}}  {10^{79}}\right)^{-1}\ ,
       \end{eqnarray}

\noindent
where  $R_{\rm univ.} \sim 1$Gpc is the comoving radius of the
universe at the reionization epoch ($(1+z)\sim 10$) and  $N_{\rm
univ.}$ is the total  number of baryons of the universe, estimated to
be $\sim 10^{79}$  baryons. Taking into account the uncertainties in
Supernova rates in the distant past, we find from
Equation~(\ref{eq:photons})  that QNe generate about
0.1-10 photons per baryon in the universe. A slightly higher (lower)
$T_{\rm pk}$ or $R_{\rm phot}$ can substantially increase (decrease)
this percentage. However,  Equation~(\ref{eq:photons}) suggests the
definite possibility that QNe can be an important source of
reionizing photons.

 \subsection{The Thomson Optical Depth}

 The Thomson optical depth is the observational link with the ionization
history of the universe.  It is defined by the relation 

 \begin{equation}\label{eq:depth} \tau_e = \int n_e(z) \sigma_T c dt =
\int f(z) n_{\rm H}(z) \sigma_T c  \frac{dz}{(1+z)H(z)}\ ,
 \end{equation}

\noindent
 where $n_e(z)$ is the co-moving electron density and
$\sigma_T=6.6524\times 10^{-25}$ cm$^2$ is the Thomson cross-section.
In the second part of Equation~(\ref{eq:depth}),  $f(z)$ is the
hydrogen ionization fraction, $n_H$ is the number density of Hydrogen,
and $H(z)$=$H_0 \left ( \Omega_{0M} (1+z)^3
+\Omega_{\Lambda}\right)^{1/2}$ is the Hubble parameter in terms of
the cosmological redshift in a  flat universe with a cosmological
constant.  We calculate $f(z)$ for QN, and then show 
how observational constraints on the optical depth provide constraints
on QN associated with Pop III stars.

Each dsQN will set up an ionization front that
propagates outward from $R_{\rm phot}(t)$. Detailed balance implies
that  the number of hydrogen atoms that are reionized $x(z)n_H(z)$ can
be determined from (see \S 2.1 in Osterbrock 1988)

  \begin{equation}\label{eq:ionrate} \left(1-x(z)\right) n_{\rm
H}\int_{13.6\ {\rm eV}}^{\infty} 4\pi j_{\gamma}  \sigma_{\rm E} dE =
x(z)^2 n_{\rm H}^2 \alpha(T)\ ,
  \end{equation}

\noindent
where $j_{\gamma}$ is the photon flux in units of cm$^{-2}$ s$^{-1}$
sr$^{-1}$  erg$^{-1}$, $\alpha(T)= 4.18\times 10^{-13} (T/10^4\ {\rm
K})^{-0.726}$  cm$^{3}$ s$^{-1}$ is the recombination coefficient 
(Mapelli \& Ferrara (2005)) and
$\sigma_{\rm E} \approx 6\times 10^{-18}$ cm$^{-2}$ is the
photo-ionization cross-section of  hydrogen atoms near threshold.

The flux $j_{\gamma}$ can be related to $N_{\rm ion}$ at a particular
distance from the QN. Since the thickness of the ionization
boundary is much smaller than the Str\"omgren sphere, we can take the
radius of the Str\"omgren sphere $R_S$ as the typical distance at
which Eq.~(\ref{eq:ionrate}) applies.  Using a typical parameter set
$R_S\approx 100$pc, $R_{\rm phot}(t)$=3$\times 10^{15}$cm and $T_{\rm
pk}$=$3\times 10^4$K, we then obtain the solution for $x(z)$ from
Eq.~(\ref{eq:ionrate}). To determine the volume fraction $f(z)$ of
ionized  hydrogen in the universe as a function of $z$, we use
$f(z)\approx  \nu(z)x(z)$ where the filling factor of non-overlapping
Str\"omgren spheres  from QNe, $\nu(z)$, is determined from (Barkana \& 
Loeb (2001)

\begin{equation}
\label{eq:hfront} \frac{d\nu(z)}{dz}=\frac{0.1x_{\rm QN}N_{I/B}}{0.76}
\left(\frac{dF_{\rm col}(z)}{dz}\right)- \frac{\alpha(T)Cn_{\rm
H}^{(0)}\nu(z)}{a^3(1+z)H(z)}
\end{equation}

\noindent
In this equation, $N_{I/B}$ denotes the number of ionizing photons released in a
QN per baryon in the star, $n_{\rm H}^{(0)}$ is the present-day
density of neutral hydrogen and $a(t)$ is the scale factor defined such
that $a(0)=1$. The factor of 0.76 in the denominator is the primordial mass 
fraction of H (we adopt 0.24 for He; see \S \ref{sec:he}). $F_{\rm col}(z)$ is the collapse fraction at high
redshift which is estimated simply as the mass fraction in halos above
the atomic cooling threshold in the Press-Schechter model (Press\&Schechter 1974).  
An order of magnitude estimate, assuming $10^{57}$ baryons in a $1.4M_{\odot}$ neutron
star gives $N_{I/B}=1000$ and yields $0.1x_{\rm QN}N_{I/B}$=10, which we keep fixed for all the curves in the 
figures below.

\begin{figure}[h!]  \vskip 0.5cm \plotone{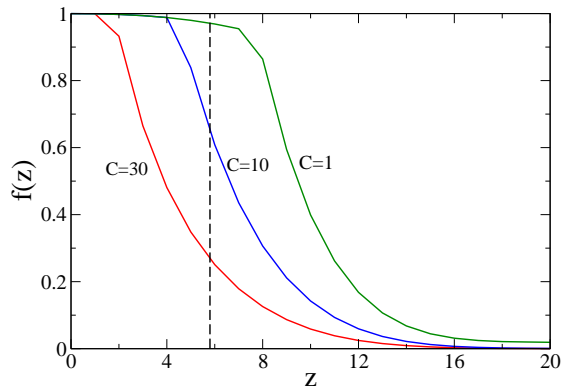}
\caption{The fraction of ionized volume in the universe $f(z)$ from
Quark-Nova events for various clumping factors $C$=1,10,30. The vertical
dashed line denotes the observed lower limit of reionization at a redshift
of $z=5.8$.}
\label{fig1}
\end{figure}

The most important factor in determining the ionization state is clumping factor $C$.
We choose $C$=1,10,30 in 
displaying results in Fig.~\ref{fig1}. These values are typical of the 
reionization era and are supported by more detailed modelling of the
evolution of the inhomogeneities of the IGM (Furlanetto \& Oh (2008a\&b), Trac \& Cen (2007), Miralda-Escud\'e et al. (2000)).

Our approximations also assume that a particular
ionization front can be associated to a unique source, which is valid
until the late stages of reionization when these fronts overlap. Since
we do not include such effects, we simply set $\nu(z)$=1 when $\nu(z)$
crosses 1, leading to the artefact of discontinuities at low redshift
in Fig.~\ref{fig1}.

The vertical dashed line shows the observational lower limit on the
reionization redshift (Fan et al. 2003) $z$=5.8. For $C\leq 10$
(moderate recombination rate), it is clear that QNe can make
a large contribution to reionization. 

\begin{figure}[h!]  \vskip 0.5cm \plotone{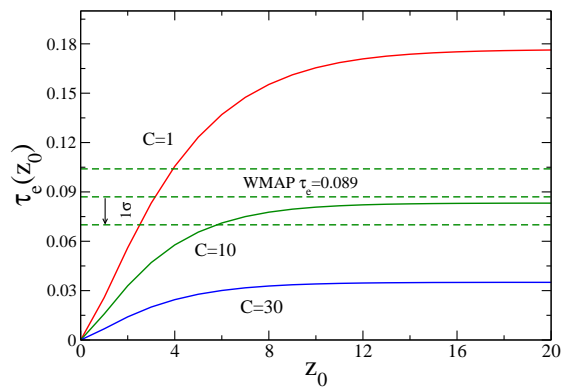}
\caption{The Thompson optical depth $\tau_e(z)$ from eq. (\ref{eq:depth})
for various clumping factors $C$=1,10,30.}
\label{fig2}
\end{figure}

We can now plot the 
results for the optical depth $\tau_e$ as a function of 
reionizing source turn-on at $z=z_0$ - shown in Fig.~\ref{fig2}.
These plots are determined from
Eq.~(\ref{eq:depth}) for $C=1,10,30$ with $f_{\rm re}$=10. The
best-fit value for $\tau_e$ from $\Lambda$CDM cosmology 
(Dunkley et al. (2009)), assuming {\it prompt} reionization at $z=z_0$, along
with $1\sigma$ limits are also shown. 

The
contribution of QNe to the optical depth depends sensitively on the
clumping factor\footnote{We have varied the clumping factor from 5-15 and find that correspondingly, $\tau_e$ varies from $\approx 0.13$ to $\approx 0.066$, assuming prompt reionization at $z\sim 20$. Thus, we do find sensitivity to the clumping factor, but this senstivity diminishes at large values of the clumping factor. A similar observation is made in relation to reionization simulations in inhomegeneous IGM (Miralda-Escud\'e, et al. 2000), where the effect of increasing overdensity saturates since late-time reionization (which is the era where the optical depth builds up) proceeds preferentially in underdense regions.}, but can be significant for $C\sim 10$, while for
$C\sim 30$ or higher, the contribution is small. If $C\sim 1$, including
QNe leads to too large an optical depth in comparison to the 5-year WMAP
data. Other reionizing sources such as Pop III stars have been studied for
consistency with the WMAP optical depth, with most reasonable models
of their evolution suggesting a ``shortfall'' in  the optical depth
($\tau_e\sim 0.05$) due to feedback effects that lead to self-regulation 
(Sokasian et al 2004), even in the case of high escape efficiency of the 
ionizing photons. Based on the results displayed in Figs.\ref{fig1} and
\ref{fig2}, we suggest the intriguing possibility that QNe can
help make up this shortfall.
  
\subsection{Helium reionization}
\label{sec:he}

Helium reionization occured more recently, as suggested by a  rapidly
varying HeII optical depth at $z\approx 2.9$ with a small line of
sight variation $\Delta z\approx 0.1$ (Reimers et al.
2005,2006). Observations of the HeII Ly$\alpha$  forest (Fardal et
al. 1998) along lines of sight to  bright quasars at $z\approx 3$ as
well as empirical  modelling of quasar luminosities and distributions
(Furlanetto \& Oh 2008a\&b)  hint that quasars are good candiates
for HeII reionization. However,  the latter work also cautions against
a simple picture in which quasars are the only HeII ionizing
source. This is because even a small fraction of HeII can absorb the
redshifted Ly$\alpha$ photons, so that a modest contribution from a
widespread low-intensity source  above the HeII edge can also explain
the apparent patchy nature of HeII  reionization at $z\gtrsim
3$. Naturally, this raises the question:  can QNe play a role
in HeliumII reionization?

First, we consider the case for HeI, which has an ionization potential
of  24.6eV. We find from Eq.(\ref{eq:ionizephotons}) that only 5\% of
the photons that ionize HI can also ionize HeI; however there are far
fewer HeI atoms $n_{\rm He}/n_{\rm H}\sim 0.08$. Since the
recombination  rates of HI and HeI differ only slightly, the
ionization front for HeI may end up closely following that of HI. If we
assume that Helium reionization does not affect HI
reionization, then for $T_{\rm pk}=3\times 10^4$K, we find that the radius 
ratio of Str\"omgren spheres $R_S^{\rm HeII}/R_S^{\rm HII}\approx 1$. 

In 
reality, the two ionization fronts are coupled since
photons above  26.4eV can ionize both H and He. While we have not performed
a detailed solution of the coupled fronts here, the results of such an 
exercise can be mocked up by a parameter 
$y=n_{{\rm HI}}a_{\nu}({\rm HI})/(n_{{\rm HI}}a_{\nu}({\rm HI})+n_{{{\rm HeI}}}a_{\nu}({{\rm HeI}}))$ which is the fraction of photons with 
energy $E=h\nu\geq 24.6$eV that are used up to ionize Hydrogen 
($n$ is the number density and $a_{\nu}$ is the photoionization cross-section). Although $a_{\nu}$ is strongly energy dependent, the 
ratio $a_{\nu}({\rm HeI})/a_{\nu}({\rm HI})$ is about 8 for 
$h\nu\gtrsim 24.6$eV. With $n_{{\rm HeI}}/n_{{\rm HI}}\sim 0.08$, we 
find $y\sim 0.65$. This implies that $(1-y)\sim 0.35$ is the fraction of 
such photons that ionize Helium. The ratio of HII to HeII Stromgren radii 
depends on $y$ as well as on the emissivity of the source (and recombination 
coefficients) and the ratio 
$R_S^{HeII}/R_S^{HII}\approx 0.35$ for $T_{\rm pk}\approx 3\times 10^4$K. 
For a higher $T_{\rm pk}\approx 4\times 10^4$K, this ratio is almost 1. 
It follows that the ratio of volume fractions $V_S^{HeII}/V_S^{HII}\approx 0.04$ for $T_{\rm pk}\approx 3\times 10^4$K and $\approx$ 1 for $T_{\rm pk}\approx 4\times 10^4$K. Including other smaller effects in Eq.(8) of our paper, 
such as the difference in recombination rates between H and He, and 
the number of ionizing photons/baryon for H and He, we find that the 
volume fraction of ionized Helium can vary dramatically from about 2\% to 
100\%  that of ionized Hydrogen. With such a strong dependence on the 
temperature of the ionizing source, it is not possible to make a robust 
claim on the efficacy of the Quark-nova on HeI reionization. 

For HeII ionization, with a large ionization threshold of 54.6 eV, the
emissivity of QNe implies that only about 1 in $10^5$ photons
can reionize HeII. Employing the front-decoupling approximation as for
HeI ionization, we find that $R_S^{\rm HeIII}/R_S^{\rm HII}\approx
0.015$ and that even for the largest $T_{\rm pk}$=$4\times 10^4$K,
approximately 5-10\% of HeII is reionized by QNe by $z\sim 3$,
depending on $C$~\footnote{This number will be lower once coupling between the HII,
HeII and HeIII fronts is introduced}. Thus, we can conservatively say that 
QNe cannot be major sources of Helium reionization but they may constitute 
a low-intensity source compared to bright Quasars.

   \section{Quark-Novae and IGM Enrichment}
     
The large value of $\tau_e$ implies that the first stars appeared as
early as $z\sim 20$, affecting the chemical evolution of the Galaxy
and subsequent star-formation. A fingerprint of early  chemical
abundances has been preserved in the extremely metal-poor stars
([Fe/H] is used here as a proxy for the metallicity) such as
HE-0107-5240 with  [Fe/H]=-5.3 (Christlieb et al. 2004, Bessell et
al. 2004) and CS 22949-037, with [Fe/H]=-4 (Depagne et al. 2002,
Israelian et al. 2004) which constrains correlations between
nucleosynthetic yields and the reionizing photon flux, since both are
affected (the former somewhat weakly) by the star's metallicity.
Previous works (Daigne et al. 2004; Tumlinson  al. 2004) concluded
that a top-heavy (40-100$M_{\odot}$) IMF with a lifetime of 50-100 Myr
is required to simultaenously satisfy constraints set by the
nucleosynthetic pattern in EMPs and early reionization. This
requirement can change once the contribution of QNe to reionization is
included. At present, we lack a comprehensive model of chemical
evolution that takes into account the contribution from QNe. However,
we can assess their importance in a global sense.

If most of the metals produced in supernovae explosions  of the first
stars are expelled into the IGM, Ricotti \& Ostriker (2004)  have
estimated that the IGM metallicity is given by

\begin{equation}
\label{eq:metals} Z_{\rm IGM}\sim (1-f_{\rm
BH})3g\times10^{-3}Z_{\odot}\left(\frac{\tau_e}{0.1}\right)
\end{equation}

\noindent
where $f_{\rm BH}$ is the collapse-fraction of massive stars into
black holes, $0.3<g<2$ depends weakly on the metallicity, and
$Z_{\odot}$  is the solar metallicity. This assumes that the  first
stars are the sole reionizing sources, in which case, one needs to
invoke a large black hole collapse fraction $\sim 0.3$ in order to
satisfy the  observational constraint on the metallicity of the early
IGM  $Z_{\rm IGM}\simeq 2\times10^{-4}Z_{\odot}$ (Schaye et
al.(2003)).  Such a large collapse fraction requires a top-heavy IMF.

QNe can change this scenario dramatically. As already shown,
QNe can generate enough photons to provide a large optical
depth, particularly if the clumping factor is small.  Consistency with
the optical depth then implies that the first stars contribute much
less to the UV photon flux, so that $\tau_e$ from the first stars can
be much  less than 0.1. From Eq.(\ref{eq:metals}), it is apparent that
even with a  small collapse fraction  into black holes and a normal
IMF, one may be able  to satisfy the observational constraint on the
IGM metallicity. It may not be necessary to invoke a large population
of short-lived massive stars very early on and the IMF of Pop III
stars could actually be more like the present-day IMF. We also note
that since QNe only produce  elements above $A\sim 130$, they
essentially do not contribute to metallicity, which is associated with
the production of much lighter elements such as C and O. Early IGM
metallicity and  reionization are decoupled in the QN
scenario. 

However, there is a link between early r-process abundance
and reionization. The amount of  r-process material ejected in a QN,
as estimated from the energetics  of the underlying phase transition,
is about $10^{-4}M_{\odot}$.   The frequency of such events during the
reionization era is   $f_{\rm dsQN}\sim .01$~yr$^{-1}$~Galaxy$^{-1}$,
so that the total  r-process material ejected by $z\sim 6$ ($10^9$
yrs) is about  $10^3M_{\odot}$.  Since the total r-process abundance
in our Galaxy  at present is  constrained to be $\sim 10^4M_{\odot}$
(Wallerstein et  al. (1997)), we conclude that about 10\% of the total
r-process elements  at  $A\gtrsim 130$ was produced early on in
QNe, with the rest   produced in Type-II supernovae or binary
neutron star mergers, as  QNe became much less frequent. In
this way, QNe also  provide a prompt and local initial
enrichment of heavy elements  $A\gtrsim 130$ that is seen in some of
the oldest stars found to  date.~\footnote{The large ratios of
[C/Fe],[O/Fe],[N/Fe] seen in the  EMPs is believed to be related to
lower mass stars, $20<M/M_{\odot}<40$  and would not be affected by
QN events, which require heavier  progenitors and only produce
elements beyond $A\sim 130$.} We plan to study the implications
for the observed scatter of r-process abundance in metal-poor stars in a 
subsequent study with a chemical evolution model.

\section{Conclusions and discussion}

We have examined the role of Quark-Novae in reionizing
the universe and contributing to its metal enrichment. 
In particular, we have suggested that:

(i) the most massive stars ($M\ge
40M_{\odot}$) in a conventional or slightly top-heavy IMF collapse to
black-holes with a possible (small) contribution to reionization from
accretion but not to metallicity;

(ii) the
reionization is driven by intermediate-mass Pop III stars,  whose higher
mass members ($25M_{\odot}\le M\le 40M_{\odot}$) end up as
QNe, providing the bulk of reionzing photons and enriching
their environment in elements beyond $A\sim 130$; and 

(iii) low
mass Pop III stars ($8M_{\odot}\le M\le 25M_{\odot}$) end up
as type II SNe. A long-lived and metal-poor population of low-mass stars 
begins to emerge at the end of the  reionization epoch (Greif et al. (2008)). 
In our scenario, the dying out of the first heavy stars coincides with a 
peak in the QN rate and therefore a peak in ionizing radiation.  

Our calculations have shown that the photon flux produced in
dual-shock Quark-Novae can be large enough to overcome recombination
in the reionization epoch. Complete reionization by $z\sim 6$ is
achieved if  the clumping factor of matter is small $C\sim{\cal
O}(1)$ but the optical depth is then $\tau_e\sim 0.17$, almost twice
that of the latest WMAP measurement . We have made estimates for 
the evolution of the ionized fraction of the IGM in a simple  model of 
non-overlapping outward-propagating ionization spheres. Our results 
imply that the optical depth from Thomson rescattering of CMB photons 
can be close to the value measured by WMAP, if the clumping factor 
$C\sim 10$. In this case, QNe provide about 60\% of the reionizing photons.
Alternatively, if $C\sim{\cal O}(30)$ or larger, QNe
play a minor role in reionization compared to the first
stars. QNe can be important for HI and HeI reionization, but
their spectra do not contain sufficient high-energy photons to
reionize HeII in any significant amount. Of central observational
interest is the production of r-process elements beyond $A\sim 130$ in
QNe, a feature which distinguishes QNe from SNe (the latter
require larger entropy than predicted in current simulations to produce
r-process elements beyond $A\sim 130$).

There are several novel aspects of the dsQN which
must be addressed in future. The most important is the final
composition of the Supernova ejecta from the first explosion which is
subsequently shocked by r-process rich QN ejecta.  This is a
critical input to chemical evolution studies which can assess the
effect of QNe on early r-process abundance patterns observed
in metal-poor stars.  In addition, the effect of
QNe on the extra-galactic radiation field as well as its
potentially large impact on the dissociation of molecular Hydrogen,
which is a ``negative feedback'' to the formation of massive stars,
remains to be analyzed. Further details relevant to the reionization era,
such as the metallicity, the evolution of the clumping factor, 
the star formation efficiency in Galaxies, and the early supernovae rate 
will have some quantitative impact on our results. These quantities are 
poorly known at present and present a source of uncertainty in any model
of reionization.

We close by predicting some important distinguishing features of
QNe that are amenable to cosmological observations:

i) \underline{Quark-Novae\ and\ nucleosynthesis}: QNe are  a
novel nucleosynthetic site for the $r$-process.  We expect the
QN ejecta to achieve $\gamma$-ray transparency  sooner than
Supernova ejecta since QN progenitors (i.e., neutron stars)
lack extended atmospheres.  Thus r-process-only nuclei with
$\gamma$-decay lifetimes of the order of years (such as $^{137}$Cs,
$^{144}$Ce, $^{155}$Eu and $^{194}$Os) can be used as tags for the
QN (Jaikumar et  al.(2007)), differentiating them from
pair-instability SNe (due to the lack of any neutron excess) or
core-collapse SNe (lower neutron excess than the QN ejecta).
This could be observed by Gamma-ray satellites such as INTEGRAL.

 ii) \underline{Quark-Novae\ and\ high\ redshift\ gamma-ray\ bursts}:   
It has been
argued that QNe  provides  an additional energy reservoir in the
context of Gamma Ray Bursts (GRBs). The QN leads to a  three stage
model for long GRBs, involving a neutron star  phase, followed by a Quark-
Star (QS) phase and a plausible third stage that occurs when the QS accretes
enough material to become a black hole.  As shown in Ouyed et al. (2007) and Staff et
al. (2007) by including the QS phase, one can account for both energy
and extended duration of the prompt emission, X-ray flaring and  the
flattening observed in GRBs light curves.  Our findings in this paper
link QNe to the reionization epoch.  The connection between QNe and
GRBs suggested above would imply that GRBs should be observed as far
back as the epoch of reionization.

To further pursue this connection between early QN and high redshift gamma ray 
sources, we note that the  recent detection of  GRB 080913 with {\it Swift} (Gehrels et
al. 2004) at redshift 6.7  makes it the highest redshift GRB to date
- more distant than the highest-redshift QSO (Fynbo et al. 2008).
At $z=6.7$ the burst occurred when the universe was less than  a Gyr
old when a high fraction of massive stars is expected to be of Pop
III.  It is  thus possible  that this   GRB  is  a member of the
long-soft GRBs produced by the collapse of the massive Pop III star
(Greiner et al. 2009).    However,   making GRBs by the collapsar
mechanism from Pop III has been questioned  since simulations  suggest
that the exploding stars will retain its hydrogen and helium rich
outer layers (e.g. Lawlor et al. 2008 and references therein).  This
violates the main requirement in the  collapsar scenario that the star
loses its extended envelope to enable the relativistic jet to punch
through the compact core on timescale for long-soft GRBs (Fryer et
al. 2001).   Pop III binary evolution can both help remove the
hydrogen envelope and spin up binary components.  However, recent simulations
question the efficiency in producing GRBs in this process
(e.g. Belczynski et al. 2007 and references therein). Furthermore,
classifying it as a long duration GRB   begs the question, in the
collapsar scenario, of  how can a massive star produce a burst as
short as 1 second?    If instead  GRB 080913 belongs to the short-hard
class (Pal'shin et al. 2008; Xu 2008), then  a neutron star-black hole
merger is favored over  a double neutron star merger, with the
Blandford-Znajek process  at play (P\'erez-Ram\'irez et al. 2008).

   The present detection rate of  GRBs at $z>5$ is about what is
predicted on the basis of the star formation rate (Jakobsson et
al. 2005). In addition, with hints of first stars having formed as
early as $z > 20$  (Kogut et al. 2003), GRBs are believed to exist as
early as $z\sim 15$-$20$.   The lack of many high-$z$ GRBs can be 
explained in our model, as a consequence of clustering of GRB events
near the end of the cosmic reionization era.  We recall  that in our
model,  the dying out of the first stars (at $z\sim 7$-$8$ when
adopting a normal IMF)  coincides with a peak in the QN rate.  This
peak in the QNe rate  (i.e.  a peak in the  GRB rate at $z\sim 7$-$8$
in our model)  offers a possible explanation for the sparsity of GRBs
out to the highest redshifts ($z > 10$) despite the immense luminosity
of both the prompt gamma-ray emission and X-ray and optical afterglows
of GRBs, and the current technology (i.e.  the high detection rate
delivered by {\it Swift}).  While a merger origin of GRB080913 cannot
be definitively ruled out, we suggest that this burst could instead
be the first  ever detection of a QN near the end of the  cosmic
reionization era.

\begin{acknowledgements}
The authors thank Ken Nollett for helpful discussions and an anonymous 
referee for useful points that improved the manuscript.  RO and REP
are supported by grants from the Natural Sciences and Engineering
Research Council  of Canada (NSERC). RO also thanks the 
Origins Institute at McMaster for hospitality and support through a Senior Visiting
Fellowship.  PJ acknowledges support from the
Department of Atomic Energy of the Government of India and from the
Argonne National Laboratory as a visiting scientist under US
Department of Energy, Office of Nuclear Physics, contract
no. DE-AC02-06CH11357.
\end{acknowledgements}

\appendix

\section{A: Pollution by CFL strangelets}
\label{appendixA}

By assuming that strange quark matter is absolutely stable, the universe might be polluted by strangelets (e.g. ejected during coalescence of quark stars in binary systems; e.g. Madsen 2005).
With a NS-NS binary collision rate of $10^{-4}-10^{-6}$yr$^{-1}$Galaxy$^{-1}$ (Belczynski et al., 2002), we expect the rate of QS-NS collisions to be at most $10^{-5}-10^{-7}$yr$^{-1}$Galaxy$^{-1}$. This will be somewhat reduced by the fact that, just as about 50\% of binary systems are disrupted by a supernova, binary neutron stars can become gravitationally unbound in a quark-nova. The amount of ejected mass will also be smaller since quark matter is stiffer than neutron matter. Following Madsen (2005), we have estimated the flux of CFL strangelets in the Earth's vicinity, accounting for factors such as the geomagnetic cutoff on the rigidity of the strangelet, and the confinement time of typical CFL strangelets in our Galaxy. We find this flux (assuming strangelet number distribution to peak in the baryon number range $A\sim 10^2-10^3$) to be $\sim 10^2-10^4$m$^{-2}$yr$^{-1}$sr$^{-1}$. This flux is too low to be tested conclusively by terrestrial experiments (although for particular values of the strangelet charge, 1 or 2 candidate events have been identified; see Finch 2006). The satellite-based AMS-02 detector launched recently has a threshold senstivity that is sufficient to detect the expected flux of CFL strangelets from quark star collisions across a wide range in $A$; these results are expected within a few years. Lunar searches, which benefit from lack of geological mixing and no magnetic field deflection, are also just approaching the required flux senstivity. Thus, at present, the strange quark matter hypothesis is not inconsistent with observations, even taking into account the binary QS-NS collisions and consequent pollution by CFL strangelets.
The most recent work about strange star binary mergers by Bauswein et. al. (2008)  found that  for high values of the MIT bag constant, strange stars could co-exist with ordinary neutron stars as they are not converted by the capture of cosmic ray strangelets. Combining their simulations  with recent estimates of stellar binary populations, Bauswein et al. (2008)  conclude that an unambiguous detection of an ordinary neutron star would  not rule out the strange matter hypothesis.

\section{B: Quark-Novae progenitors}
\label{appendixB}

The fit to the observed light curve of SN2006gy, Leahy \& Ouyed (2008) assumes
 QNe progenitor mass in the (40-60)$M_{\odot}$ range.
 However, one can employ the parameter degeneracy in that fit to examine the dual-shock scenario with the more conservative mass range of (25-40)$M_{\odot}$ which is more in line with the literature  (e.g. Heger et al. ApJ 591 2003; Nakazato et al. 2008) which suggets prompt BH formation above 
 $40M_{\odot}$. It should be noted however that the effect of the fireball in the CFL phase 
 (Ouyed et al. 2005) has not been taken into account in any of these simulations, which means the range 25-40$M_{\odot}$ could be somewhat underestimated.

\section{C: Shock efficiency in dual-shock Quark-Novae}
\label{appendixC}
  
  According to Ouyed et al. (2007), the shock efficiency varies as $\rho_{\rm env.}^2$ with the mean
 SN envelope density  given by $\rho_{\rm env}\propto M_{\rm env.}/R_{\rm env}^3$. If we choose
40$M_{\odot}$ instead of 60$M_{\odot}$ for the progenitor of the SN,
and demand the same efficiency, we find that
the collision radius should be $R_{40} =
(40/60)^{1/3}\times R_{60}\sim 0.876\times R_{60}$ and the delay time $t_{\rm delay, 40} = 0.876 t_{\rm delay, 60}\sim 0.876\times 15$ days $\sim 13$ days. Similarly, $T_{\rm pk, 40} = (R_{40}/R_{60})^{1/2}T_{\rm pk, 60} = (0.876)^{1/2}T_{\rm pk, 60}\sim 0.94T_{\rm pk, 60}$. With these changes, we find a reduction factor of 0.6
in the total number of ionizing photons. This does not change our order of magnitude arguments on the ionization efficiency and we had in any case adopted a very conservative estimate for the number of ionizing photons initially by neglecting photons emitted after the ejecta becomes transparent.

\end{document}